\documentclass[aps,prd,amsmath,superscriptaddress,twocolumn,floatfix,nofootinbib,preprintnumbers]{revtex4-2}
\usepackage{amssymb}\usepackage{txfonts}\usepackage{epsfig}\usepackage{bm}
\usepackage{color}\usepackage{graphicx,graphics}\usepackage{multirow}\usepackage{float}\usepackage{ulem}
\usepackage{setspace}\usepackage{slashed}\usepackage{booktabs}\usepackage{hyperref}
\hypersetup{colorlinks=true, citecolor=blue, linkcolor=red, filecolor=black,urlcolor=blue}
\allowdisplaybreaks[4]

\begin{document}

\title{Transverse momentum asymmetry in the semi-inclusive electron positron \\ annihilation process}

\author{Weihua Yang}

\affiliation{College of Nuclear Equipment and Nuclear Engineering, Yantai University,\\ Yantai, Shandong 264005, China}

\author{Xing-hua Yang}

\affiliation{School of Physics and Optoelectronic Engineering, Shandong University of Technology,\\ Zibo, Shandong 255000, China}


\author{Zhe Zhang}

\affiliation{Southern Center for Nuclear-Science Theory (SCNT), Institute of Modern Physics, Chinese Academy of Sciences, Huizhou 516000, China}

\author{Jing Zhao}\thanks{Corresponding author:zhaojingzj@sdu.edu.cn}

\affiliation{Key Laboratory of Particle Physics and Particle Irradiation (MOE), Institute of Frontier and Interdisciplinary Science, Shandong University, \\Qingdao, Shandong 266237, China}

\begin{abstract}

Hadronization, a nonperturbative process, cannot be calculated from first principles. It can be investigated either by using phenomenological models or by examining the behavior of produced hadrons or through fragmentation functions. These fragmentation functions are nonperturbative quantities whose determination relies entirely on experimental data. However, higher-twist fragmentation functions present significant challenges for their determination due to power suppression.
In this paper, we propose an asymmetry to study twist-3 fragmentation functions. This asymmetry is defined as the transverse momentum asymmetry of the fragmenting quark and/or the produced jet with respect to the observed hadron direction within the semi-inclusive electron positron annihilation process. As a twist-3 effect, this asymmetry is sensitive to the distribution of the jet relative to the produced hadron direction during hadronization. Furthermore, it is closely related to twist-3 transverse momentum dependent fragmentation functions and provides a set of measurable quantities for their determination.

\end{abstract}

\maketitle

\section{Introduction}

Quantum chromodynamics (QCD) is a fundamental non-Abelian gauge theory of strong interaction whose Lagrangian deals with quarks, gluons, and their interactions. It is characterized by two key features: asymptotic freedom and color confinement. Due to color confinement, isolated quarks and gluons are not experimentally observable; only color-singlet hadrons can be detected. The process in which free quarks and/or gluons combine to form color singlet hadrons is known as hadronization. Hadronization is inherently nonperturbative and cannot be calculated from first principles. Nevertheless, it can be described by phenomenological models, such as the Feynman-Field model \cite{Field:1977fa}, the LUND model \cite{Andersson:1983ia}, the Webber model \cite{Marchesini:1983bm,Webber:1983if}, and the recombination model \cite{Anisovich:1972pq,Bjorken:1973mh,Das:1977cp,Xie:1988wi}. While these models provide a global perspective on hadronization, they often lack detailed insights. To address this, one can study the behaviors of the single hadron or the double-hadron during hadronization process. They respectively correspond to the single-hadron fragmentation functions (FFs) \cite{Berman:1971xz,Collins:1981uw} and the double-hadron fragmentation functions (diFFs) \cite{Bianconi:1999cd,Bianconi:1999uc,Bacchetta:2003vn}.
Both FFs and diFFs are nonperturbative quantities; a systematic introduction to them can be found in Ref. \cite{Metz:2016swz}. 

In addition to understanding behaviors of the single hadron or the double hadron, it is also very important to study the collective behavior of multiple hadrons during the hadronization process. An ensemble of hadrons clustered within a phase space defined by specific parameters constitutes a jet. Jets originating from quarks (or gluons) have received increasing attention in recent years as effective probes of hadronization. Theoretical formulations for jet production in high-energy collisions exhibit simpler forms and avoid introducing additional uncertainties associated with (di)FFs. This is helpful to improve the accuracy of measurements. 
Theoretical studies of jet production in high-energy collisions encompass two approaches: one focuses on higher-twist calculations up to leading order \cite{Song:2010pf,Song:2013sja,Wei:2016far,Yang:2017sxz,Chen:2020ugq,Yang:2020qsk,Yang:2022knp,Yang:2022xwy,Yang:2023vyv,Yang:2023zod,Yang:2025hky}, while the other focuses on higher-order calculations \cite{Gutierrez-Reyes:2018qez,Gutierrez-Reyes:2019vbx,Liu:2018trl,Liu:2020dct,Kang:2020fka,Arratia:2020ssx,Arratia:2019vju,Arratia:2022oxd}.
Furthermore, the introduction of jet fragmentation functions (JFFs) \cite{Procura:2009vm,Jain:2011xz,Jain:2011iu,Chien:2015ctp,Arleo:2013tya,Kaufmann:2015hma,Kang:2016ehg,Dai:2016hzf,Kang:2019ahe,Kang:2017glf,Kang:2020xyq}, designed to study hadron distributions within a jet, has significantly broadened the research area. An observed hadron carries not only a longitudinal momentum fraction relative to the jet but also a transverse momentum relative to the jet axis. Consequently, JFFs encode a set of angular modulations of the hadron distributions within the jet, providing valuable insights into the hadronization process. Conversely, in the hadron collinear frame, the jet axis acquires transverse momentum ($\vec{j}_T$) relative to the direction of the observed hadron. Under the approximation that the jet axis aligns with the direction of the initial parton (quark or antiquark), the intrinsic transverse momentum of the parton satisfies $\vec{k}^\prime_T = \vec{j}_T$. Therefore, unlike JFFs, which probe the distributions of a hadron within the jet, $\vec{j}_T$ serves as an effective probe for studying jet distributions relative to the observed hadron direction.

To investigate the transverse momentum effects associated with the jet during hadronization, we provide a general theoretical framework for the jet production semi-inclusive electron positron annihilation process in this paper. Semi-inclusive implies a back-to-back jet is also measured in addition to the observed hadron. Our method is consistent with that of Refs. \cite{Yang:2017sxz,Yang:2022knp}, which focused on leading-order higher-twist calculations. This approach is motivated by the fact that transverse momenta typically manifest as higher-twist effects relative to the momentum of the observed hadron.
We first present a detailed calculation for the differential cross section up to twist-3 and express it in terms of transverse momentum dependent fragmentation functions (TMD FFs) \cite{Boer:1997mf}. Subsequently, we calculate the asymmetry, defined as the transverse momentum asymmetry of the fragmenting quark and/or the produced jet with respect to the observed hadron direction in the semi-inclusive annihilation process. We note that this asymmetry reveals intrinsic transverse momentum effects and also provides insights into jet distributions relative to the observed hadron. Crucially, this asymmetry is intimately related to twist-3 TMD FFs and offers a set of measurable quantities for their determination.

To be explicit, we organize this paper as follows. In Sec. \ref{sec:frame}, we present an introduction of the reference frame used in this paper and show the definition of the transverse momentum asymmetry in this frame. In Sec. \ref{sec:calculations}, we present the general theoretical framework for calculating the hadronic tensor and the differential cross section up to twist-3. In Sec. \ref{sec:results}, we perform the numerical estimates for the transverse momentum asymmetry. Finally, we give a brief summary in Sec. \ref{sec:summary}.

\section{Transverse momentum asymmetry}\label{sec:frame}

\begin{figure}
  \centering
 \includegraphics[width=0.8\linewidth]{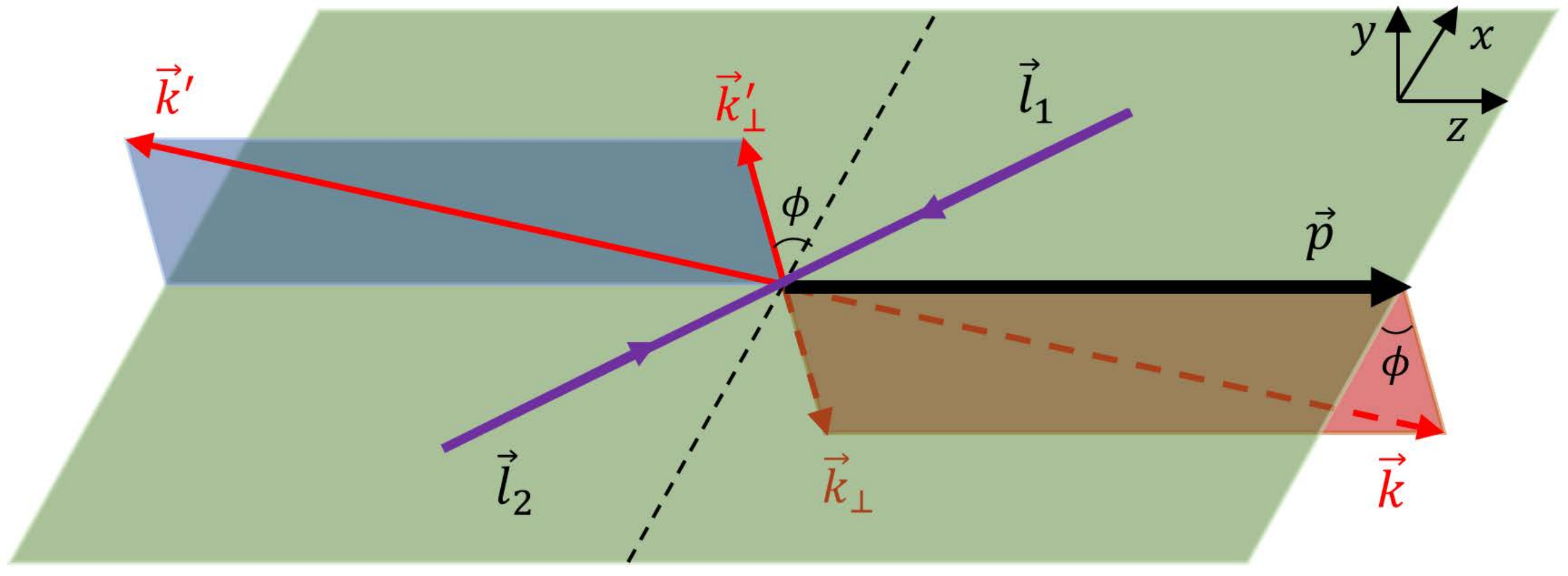}\\
  \caption{The center-of-mass frame of leptons of the semi-inclusive electron positron annihilation process.}\label{fig:jet}
\end{figure}

As mentioned in the Introduction, we consider the semi-inclusive electron positron annihilation process.
We denote this process as 
\begin{align}
    e^-(l_1) + e^+(l_2)\to h(p) + jet (k') +X,
\end{align}
where the momenta of the relevant particles are given in parentheses.
To define the transverse momentum asymmetry, we need to establish a reference frame and determine the transverse direction.
We begin by introducing the center-of-mass frame of the leptons, where the observed hadron travels along the $+z$ direction and the momenta of leptons lie in the $x$-$z$ plane, also referred to as the reaction plane (see Fig. \ref{fig:jet}). The transverse direction is then defined as the direction perpendicular to the $z$ axis, lying within the $x$-$y$ plane. 
In this frame, we can introduce the light-cone unit vectors, $\bar{n}^\mu=(1,0,\vec{0}_T)$ and $n^\mu=(0,1,\vec{0}_T)$, which satisfy $\bar{n}^2=n^2=0$ and $\bar{n}\cdot n=1$. 
These vectors allow us to define the transverse tensors for further analysis,
\begin{align}
    & g^{\mu\nu}_T=g^{\mu\nu}-\bar{n}^\mu n^\nu - \bar{n}^\nu n^\mu, \label{e.gT}\\
    & \varepsilon_T^{\mu\nu}=\varepsilon^{\mu\nu\alpha\beta}\bar{n}_\alpha n_\beta.\label{e.epsilonT}
\end{align}
The momenta of the relevant particles can be parametrized as  
\begin{align}
   & p^\mu=\frac{zQ}{\sqrt{2}}\bar{n}^\mu + \frac{M^2}{\sqrt{2}zQ}n^\mu, \\
   & q^\mu=\frac{Q}{\sqrt{2}}\bar{n}^\mu + \frac{Q}{\sqrt{2}}n^\mu , \\
   & l_1^\mu=\frac{(1-y)Q}{\sqrt{2}}\bar{n}^\mu + \frac{yQ}{\sqrt{2}} n^\mu +l_T^\mu, \\
   & l_2^\mu=\frac{yQ}{\sqrt{2}}\bar{n}^\mu + \frac{(1-y)Q}{\sqrt{2}} n^\mu -l_T^\mu. 
\end{align}
Neglecting the hadron mass $M$ yields $p^\mu \sim zQ\bar{n}^\mu/\sqrt{2}$. Here $Q^2=q^2=(l_1+l_2)^2$. The commonly used variables are defined as 
\begin{align}
    & z=\frac{2p\cdot q}{Q^2}, && y=\frac{p\cdot l_1}{p\cdot q}.
\end{align}

To decompose the leptonic tensor illustrated in the following section, we also introduce the timelike vector $\hat{t}^\mu$ and the spacelike vector $\hat{v}^\mu$, 
\begin{align}
    & \hat{t}^\mu=\frac{q^\mu}{Q}, && \hat{v}^\mu=\frac{2p^\mu}{zQ}-\hat{t}^\mu.
\end{align}
They satisfy the orthogonal relation $\hat{t}\cdot \hat{v}=0$. From the previous introduction, we notice that $(\bar{n}, n)$ and $(\hat{t}, \hat{v})$ are equivalent, 
\begin{align}
     & \hat{t}^\mu=\frac{1}{\sqrt{2}}\left(\bar{n}^\mu+n^\mu\right), \\ 
     & \hat{v}^\mu=\frac{1}{\sqrt{2}}\left(\bar{n}^\mu-n^\mu\right).  
\end{align}
Under this circumstance, the dimensionless tensors can be defined in terms of $\hat{t}$ and $\hat{v}$,
\begin{align}
    & g^{\mu\nu}_\perp=g^{\mu\nu}-\hat{t}^\mu \hat{t}^\nu + \hat{v}^\mu \hat{v}^\nu, \\
    & \varepsilon_\perp^{\mu\nu}=-\varepsilon^{\mu\nu\alpha\beta}\hat{t}_\alpha \hat{v}_\beta,
\end{align}
and they satisfy $g^{\mu\nu}_\perp= g^{\mu\nu}_T$ and $\varepsilon_\perp^{\mu\nu}=\varepsilon_T^{\mu\nu}$.

In the center-of-mass frame of the leptons illustrated in Fig. \ref{fig:jet}, the quark and anti-quark are produced back-to-back at tree level without higher-order gluon radiations (two-jet event).
Their momenta are labeled as $k$ and $k'$, respectively. In the jet production semi-inclusive annihilation considered in this paper, we take $\vec{k}^\prime$ as the momentum of the jet under the approximation that the jet axis direction coincides with that of the antiquark. The hadron is detected in the back-to-back hemisphere. 
Given that the transverse momentum is defined in the $x$-$y$ plane, the transverse momentum asymmetry can be expressed as a difference of the jet transverse momentum ($k_\perp^\prime$). 
Therefore, we have
\begin{align}
    & A^x=\frac{{k}^{\prime x}_\perp(x<0)-{k}^{\prime x}_\perp(x>0)}{{k}^{\prime x}_\perp(x<0)+{k}^{\prime x}_\perp(x>0)}, \label{f:axk} \\
    & A^y=\frac{{k}^{\prime y}_\perp(y<0)-{k}^{\prime y}_\perp(y>0)}{{k}^{\prime y}_\perp(y<0)+{k}^{\prime y}_\perp(y>0)}. \label{f:ay}
\end{align}
We can see that asymmetry $A^x (A^y)$ reveals the difference of the transverse momenta in the $x (y)$ direction. It thus provides insights into transverse momentum effects or jet distributions relative to the observed hadron direction. Since $\vec{k}_\perp^\prime=-\vec{k}_\perp$, $A^x$ and $A^y$ can be rewritten in terms of the intrinsic 
transverse momentum of the fragmenting quark ($k_\perp$)
\begin{align}
    & A^x=\frac{{k}^x_\perp(x>0)-{k}^x_\perp(x<0)}{{k}^x_\perp(x>0)+{k}^x_\perp(x<0)}, \label{f:axk} \\
    & A^y=\frac{{k}^y_\perp(y>0)-{k}^y_\perp(y<0)}{{k}^y_\perp(y>0)+{k}^y_\perp(y<0)}. \label{f:ayk}
\end{align}
The definitions in Eqs. \eqref{f:axk} and \eqref{f:ayk} provide a convenient way to study the intrinsic transverse momentum of the fragmenting quark and twist-3 TMD FFs. It is helpful for understanding the hadronization process. In the following, we only consider the definitions in Eqs. \eqref{f:axk} and \eqref{f:ayk}.

\section{Theoretical framework} \label{sec:calculations}

Precise measurements of differential cross sections are essential for investigating higher-twist effects. In this section, we present the theoretical framework for the jet production semi-inclusive electron positron annihilation process.
For completeness, we incorporate both electromagnetic interaction and weak interaction. 

The differential cross section of this process can be written in a unified form,
\begin{align}
    \frac{d\sigma}{dzdyd^2k_\perp^\prime}=\frac{\alpha^2}{Q^4}\frac{z}{4\pi}A_r L_r^{\mu\nu}(l_1, l_2) W_{\mu\nu}^r(q,k_\perp^\prime, p),
\end{align}
where $\alpha$ is the fine structure constant. The kinematic factors $A_r$ ($r=\gamma\gamma,\, \gamma Z,\, ZZ$) are given by
\begin{align}
    & A_{\gamma\gamma} =e_l^2 e_q^2, \\
    & A_{\gamma Z} =\frac{2e_l e_q Q^2(Q^2-M_Z^2)}{\left[(Q^2-M_Z^2)^2+\Gamma^2_Z M_Z^2\right]\sin^2 2\theta_W}, \\
    & A_{ZZ} =\frac{Q^4}{\left[(Q^2-M_Z^2)^2+\Gamma^2_Z M_Z^2\right]\sin^4 2\theta_W}.
\end{align}
Here, $e_l$ and $e_q$ are electric charges of relevant leptons and quarks, $M_Z$ and $\Gamma_Z$ are, respectively, the mass and width of the $Z^0$ boson, and $\theta_W$ is the weak mixing angle. 

The leptonic tensors for the electromagnetic term, the interference term, and the weak term are given by
\begin{align}
    & L_{\gamma\gamma}^{\mu\nu}(l_1, l_2)=2\left(l_1^\mu l_2^\nu + l_1^\nu l_2^\mu -g^{\mu\nu} l_1\cdot l_2\right) +2i\lambda_e \varepsilon^{\mu\nu l_1l_2}, \label{f:leptongammagamma}\\
    & L_{\gamma Z}^{\mu\nu}(l_1, l_2)= \left(c_V^e-c_A^e \lambda_e\right)L_{\gamma\gamma}^{\mu\nu}(l_1, l_2), \label{f:leptongammaz}\\
    & L_{Z Z}^{\mu\nu}(l_1, l_2)= \left(c_1^e-c_3^e \lambda_e\right)L_{\gamma\gamma}^{\mu\nu}(l_1, l_2), \label{f:leptonzz}
\end{align}
where $\lambda_e$ is the helicity of the initial electron, and $c_V^e$ and $c_A^e$ are defined in the weak current $J^\mu=\bar{\psi}\gamma^\mu(c_V^e-c_A^e\gamma^5)\psi$ with $c_1^e=(c_V^e)^2+(c_A^e)^2$ and $c_3^e=2c_v^e c_A^e$. We can rewrite the leptonic tensor in terms of vectors $\hat{t}$ and $\hat{v}$ as follows:
\begin{align}
    L_{\gamma\gamma}^{\mu\nu}=&Q^2\Bigg[-A(y)g_\perp^{\mu\nu}+B(y)\hat{v}^\mu\hat{v}^\nu-B(y)\left(\hat{x}^\mu\hat{x}^\nu+\frac{1}{2}g_\perp^{\mu\nu}\right)\nonumber\\
    & - 2C(y)D(y)\hat{v}^{\{\mu}\hat{x}^{\nu\}}\Bigg]\nonumber\\
    + & i\lambda_e Q^2 \left[C(y)\varepsilon_\perp^{\mu\nu}-2D(y)\varepsilon_\perp^{\rho[\mu}\hat{v}^{\nu]}\hat{x}_\rho\right], \label{f:lepto}
\end{align}
where the unit vector $\hat{x}^\mu$ is defined as   $\hat{x}^\mu=l_T^\mu/QD(y)$.
For simplicity, the functions of $y$ are written as
\begin{align}
    & A(y)=2y^2-2y+1, \\
    & B(y)=4y(1-y),  \\
    & C(y)=1-2y, \\
    & D(y)=\sqrt{y(1-y)}.
\end{align}
The expression for the leptonic tensor in Eq. \eqref{f:lepto} can greatly simplify the contraction with the hadronic tensor in the framework of the parton model.

The $k_\perp^\prime$ dependent hadronic tensor is obtained by integrating over $k_z^\prime$,
\begin{align}
    W_{\mu\nu}^r(q,k_\perp^\prime, p)=\int \frac{dk_z^\prime}{(2\pi) 2E^\prime} W_{\mu\nu}^r(q,k^\prime, p), 
\end{align}
where $W_{\mu\nu}^{ZZ}(q,k^\prime, p)$ is defined as
\begin{align}
    W_{\mu\nu}^{ZZ}(q,k^\prime, p) & = \sum_X(2\pi)^3\delta^4(q-p-k^\prime-p_X) \nonumber \\
    &\times\langle 0|J^Z_\mu(0)|pk'X\rangle\langle pk'X|J_\nu^Z(0)|0\rangle. 
\end{align}
The hadronic tensors for the electromagnetic term, $W_{\mu\nu}^{\gamma\gamma}$, and the interference term, $W_{\mu\nu}^{\gamma Z}$, are defined similarly and can be obtained by replacing $ZZ$ with $\gamma\gamma$ and $\gamma Z$, respectively. We note that the complete results should include the electromagnetic term, the interference term, and the weak term.

\subsection{Hadronic tensor in the parton model}

The hadronic tensor contains nonperturbative information and can be calculated in terms of the structure functions \cite{Yang:2017sxz,Yang:2022mmh}. Given that our focus is on transverse momentum effects and twist-3 FFs, we do not introduce structure functions and instead restrict the analysis to calculation within the parton model framework.

At the tree level of perturbative QCD, the hadronic tensor can be expressed as 
\begin{align}
    W_{\mu\nu}(q,k_\perp^\prime)=\sum_{i,c} W^{(j,c)}_{\mu\nu}(q,k_\perp^\prime)
\end{align}
after using collinear expansion \cite{Wei:2016far}. Here $j$ and $c$ denote the number of gluons and cut positions, respectively. 
$W^{(j,c)}_{\mu\nu}$ can be expressed by a trace of the collinear-expanded hard part and gauge invariant quark-$j$-gluon-quark correlator. The explicit expressions up to twist-3 are given by
\begin{align}
   & W^{(0)}_{\mu\nu}
   = \frac{1}{2} \mathrm{Tr}\big[\hat h^{(0)}_{\mu\nu} \hat\Xi^{(0)}\big],\label{f:tW0}\\
   & W^{(1,L)}_{\mu\nu}
   =-\frac{1}{4(p\cdot q)} \mathrm{Tr}\big[\hat h^{(1)\rho}_{\mu\nu}\hat\Xi^{(1)}_{\rho} \big],\label{f:tW1L}
\end{align}
where arguments $(q,k_\perp^\prime)$ have been omitted for simplicity. The hard parts are universal and written as 
\begin{align}
&\hat h^{(0)}_{\mu\nu} =\frac{1}{p^+} \Gamma_\mu^q \slashed n \Gamma_\nu^q , \label{f:h0}\\
&\hat h_{\mu\nu}^{(1)\rho} = \Gamma_\mu^q \slashed n \gamma^\rho \slashed{\bar n}\Gamma_\nu^q,  \label{f:h1}
\end{align}
with $\Gamma_\mu^q=\gamma_\mu(c_V^q-c_A^q\gamma^5)$. The corresponding correlators are defined as  
\begin{align}
\hat \Xi^{(0)} =& \sum_X \int \frac{p^+d \xi^-d^2\vec{\xi}_\perp}{2\pi} e^{-ip^+\xi^-/z+i\vec{k}_\perp\vec{\xi}_\perp}
  \langle 0 | \mathcal{L}^\dagger(0,\infty) \nonumber\\
 &\times \psi(0) |hX\rangle\langle hX| \bar\psi(\xi^-) \mathcal{L}(\xi^-,\infty) |0\rangle, \label{f:Xi0}\\
  \hat \Xi^{(1)}_{\rho}  = & \sum_X  \int\frac{p^+d\xi^-d^2\vec{\xi}_\perp}{2\pi} e^{-ip^+\xi^-/z+i\vec{k}_\perp\vec{\xi}_\perp} \langle 0 | \mathcal{L}^\dagger (0,\infty)  \nonumber\\
&\times D_{\rho}(0) \psi(0) |hX\rangle\langle hX| \bar\psi(\xi^-) \mathcal{L}(\xi^-,\infty) |0\rangle. \label{f:t3Xi1} 
\end{align}
We note that $\mathcal{L}(0,y)$ is the gauge link while $D_\rho=i\partial_\rho-gA_\rho$ is the transverse covariant derivative.

\subsection{Fragmentation functions}

Correlators shown in Eqs. \eqref{f:Xi0} and \eqref{f:t3Xi1} are $4\times 4$ matrices that can be decomposed by Dirac matrices.
To satisfy the helicity conservation, we only need to consider the chiral-even terms, $\gamma^\alpha$, and the $\gamma^5\gamma^\alpha$ terms, when decomposing the correlators. For example, $\hat \Xi^{(0)}=\gamma^\alpha \Xi_\alpha^{(0)}+\gamma^5\gamma^\alpha \tilde\Xi_\alpha^{(0)}$. Here $\Xi_\alpha^{(0)}$ and  $\tilde \Xi_\alpha^{(0)}$ are coefficient functions, which can be obtained by 
\begin{align}
 & \Xi_\alpha^{(0)} =\frac{1}{4}\mathrm{Tr}[\gamma^\alpha \hat\Xi^{(0)}], \\
 & \tilde\Xi_\alpha^{(0)} =\frac{1}{4}\mathrm{Tr}[\gamma^\alpha\gamma^5 \hat\Xi^{(0)}].
\end{align}
Coefficient functions can be further decomposed in terms of TMD FFs. For $\hat{\Xi}^{(0)}$ we have 
\begin{align}
  z\Xi^{(0)}_\alpha = &p^+\bar{n}_\alpha \left(D_1 +\frac{\varepsilon_\perp^{kS}}{M}D^\perp_{1T}\right)-k_{\perp\alpha} D^\perp - M\varepsilon^S_{\perp ~ \alpha}D_T \nonumber \\
  &-\varepsilon^k_{\perp~\alpha}\left(\lambda_h D^\perp_L +\frac{k_\perp \cdot S_T}{M}D^\perp_T\right), \label{f:xi0even}\\
  z\tilde\Xi^{(0)}_\alpha = &-p^+\bar{n}_\alpha\left(\lambda_hG_{1L} +\frac{k_\perp \cdot S_T}{M}G^\perp_{1T}\right) -\varepsilon^k_{\perp~\alpha}G^\perp  \nonumber \\
  & + M S_{T\alpha}G_T+k_{\perp\alpha}\left(\lambda_h G^\perp_L +\frac{k_\perp \cdot S_T}{M}G^\perp_T\right). \label{f:xi0odd}
\end{align}
Here $p^+=p\cdot n$ and $\lambda_h$ is the helicity of the produced hadron, and $S_T$ is the transverse component of its polarization. The latter can be parametrized as 
\begin{align}
    S_T^\mu=|\vec{S}_T|(0, \cos\phi_S, \sin\phi_S, 0).
\end{align} 
For $\hat{\Xi}^{(1)}_\rho$ we have 
\begin{align}
  z\Xi^{(1)}_{\rho\alpha}=&-p^+\bar n_\alpha k_{\perp\rho} D_d^\perp - p^+\bar n_\alpha \Bigg[M\varepsilon^S_{\perp ~ \rho}D_{dT} \nonumber \\
  & +\varepsilon^k_{\perp~\rho}\left(\lambda_h D^\perp_{dL} +\frac{k_\perp \cdot S_T}{M}D^\perp_{dT}\right)\Bigg], \label{f:xi1even3} \\
  z\tilde\Xi^{(1)}_{\rho\alpha}=&ip^+\bar n_\alpha \varepsilon^k_{\perp~\rho} G_d^\perp - ip^+\bar n_\alpha \Bigg[MS_{T\rho}G_{dT}\nonumber \\
  & +k_{\perp\rho}\left(\lambda_h G^\perp_{dL} +\frac{k_\perp \cdot S_T}{M}G^\perp_{dT}\right)\Bigg].\label{f:xi1odd3}
\end{align}
Here subscript $d$ is used to denote TMD FFs obtained from correlator $\hat{\Xi}^{(1)}_\rho$.

\subsection{Calculations up to twist-3}

The contribution of the leading twist comes only from the quark-quark correlator $\hat \Xi^{(0)}$. To calculate it, we use Eqs. (\ref{f:xi0even}) and (\ref{f:xi0odd}), and the following relations:
\begin{align}
  & \mathrm{Tr}\big[\hat h^{(0)}_{\mu\nu} \slashed {\bar n}\big]
 =-\frac{4}{p^+}\left( c_1^qg_{\perp\mu\nu}+ic_3^q\varepsilon_{\perp\mu\nu}\right),\label{f:traceh0} \\
  & \mathrm{Tr}\big[\hat h^{(0)}_{\mu\nu} \gamma^5 \slashed {\bar n}\big]
 =\frac{4}{p^+}\left( c_3^qg_{\perp\mu\nu}+ic_1^q\varepsilon_{\perp\mu\nu}\right).\label{f:traceh05}
\end{align}
Substituting them into Eq. (\ref{f:tW0}) yields the leading twist contribution to the hadronic tensor,
\begin{align}
  z W_{t2\mu\nu} =& -2\Big[c_1^qg_{\perp\mu\nu}+ic_3^q\varepsilon_{\perp\mu\nu}\Big]\left(D_1 +\frac{\varepsilon_\perp^{kS}}{M}D^\perp_{1T}\right)\nonumber\\
 &-2\Big[c_3^qg_{\perp\mu\nu}+ic_1^q\varepsilon_{\perp\mu\nu}\Big]\left(\lambda_h D_{1L} +\frac{k_\perp \cdot S_T}{M}G^\perp_{1T}\right). \label{f:Wt2munu}
\end{align}

Nevertheless, twist-3 contributions have two sources, one is the quark-quark correlator $\hat \Xi^{(0)}$ and  the other is the quark-gluon-quark correlator $\hat \Xi^{(1)}_\rho$. We first calculate these contributions from the quark-quark correlator $\hat \Xi^{(0)}$. Here we use the following traces:
\begin{align}
  & \mathrm{Tr}\big[\hat h^{(0)}_{\mu\nu} \gamma^\alpha\big]k_{\perp\alpha}
 =\frac{4}{p^+}\left( c_1^qk_{\perp\{\mu} n_{\nu\}}-ic_3^q \varepsilon^k_{\perp[\mu} n_{\nu]} \right),\label{f:trt3h0} \\
  & \mathrm{Tr}\big[\hat h^{(0)}_{\mu\nu} \gamma^5 \gamma^\alpha\big]k_{\perp\alpha}
 =-\frac{4}{p^+}\left( c_3^qk_{\perp\{\mu} n_{\nu\}}-ic_1^q \varepsilon^k_{\perp[\mu} n_{\nu]} \right),\label{f:trt305} 
\end{align}
where $A_{\{\mu}B_{\nu\}}=A_\mu B_\nu+A_\nu B_\mu$ and $A_{[\mu}B_{\nu]}=A_\mu B_\nu-A_\nu B_\mu$.
Using Eqs. (\ref{f:xi0even}) and  (\ref{f:xi0odd}) and substituting them into Eq. (\ref{f:tW0}), we obtain
\begin{align}
  z W^{(0)}_{t3\mu\nu} = & -\frac{2}{p^+}\left[c_1^q  k_{\perp\{\mu}n_{\nu\}}-ic_3^q \varepsilon^k_{\perp[\mu}n_{\nu]}\right]D^\perp  \nonumber\\
  & -\frac{2M}{p^+}\left[c_1^q  \varepsilon^S_{\perp\{\mu}n_{\nu\}}+ic_3^q S_{T[\mu}n_{\nu]}\right]D_T  \nonumber\\
  & - \frac{2}{p^+}\left[c_1^q  \varepsilon^k_{\perp\{\mu}n_{\nu\}}+ic_3^q k_{\perp[\mu}n_{\nu]}\right]\nonumber \\
  &\quad \quad \times\left(\lambda_h D^\perp_{L} +\frac{k_\perp \cdot S_T}{M}D^\perp_{T}\right) \nonumber\\
  & +\frac{2}{p^+}\left[c_3^q  \varepsilon^k_{\perp\{\mu}n_{\nu\}}+ic_1^q k_{\perp[\mu}n_{\nu]}\right]G^\perp  \nonumber\\
  & -\frac{2M}{p^+}\left[c_3^q  S_{T\{\mu}n_{\nu\}}-ic_1^q \varepsilon^S_{\perp[\mu}n_{\nu]}\right]G_T  \nonumber\\
  & - \frac{2}{p^+}\left[c_3^q  k_{\perp\{\mu}n_{\nu\}}-ic_1^q \varepsilon^k_{\perp[\mu}n_{\nu]}\right]\nonumber \\
  &\quad \quad \times\left(\lambda_h G^\perp_{L} +\frac{k_\perp \cdot S_T}{M}G^\perp_{T}\right).
 \label{f:W0t3}
\end{align}

For the twist-3 contribution from the correlator $\hat \Xi^{(1)}_\rho$, we use the following traces:
\begin{align}
 & \mathrm{Tr}\big[\hat h^{(1)\rho}_{\mu\nu} \slashed{\bar n}\big]k_{\perp\rho}
 = -8\big(c_1^q k_{\perp\mu} \bar n_\nu -ic_3^q \varepsilon^k_{\perp\mu}\bar n_\nu \big), \label{f:trt3h1}\\
 & \mathrm{Tr}\big[\hat h^{(1)\rho}_{\mu\nu}\gamma^5 \slashed{\bar n}\big]k_{\perp\rho}=-8\big(i c_1^q \varepsilon^k_{\perp\mu}\bar n_\nu-c_3^q k_{\perp\mu} \bar n_\nu  \big), \label{f:trht3h51}
\end{align}
and obtain the following result:
\begin{align}
  z W^{(1,L)}_{t3\mu\nu} = & +\frac{2}{p\cdot q}\left[c_1^q k_{\perp\mu}\frac{p_{\nu}}{z} -ic_3^q \varepsilon^k_{\perp\mu}\frac{p_{\nu}}{z}\right]D^\perp  \nonumber\\
  & +\frac{2M}{p\cdot q}\left[c_1^q  \varepsilon^S_{\perp\mu}\frac{p_{\nu}}{z}+ic_3^q S_{T\mu}\frac{p_{\nu}}{z}\right]D_T  \nonumber\\
  & + \frac{2}{p\cdot q}\left[c_1^q  \varepsilon^k_{\perp\mu}\frac{p_{\nu}}{z}+ic_3^q k_{\perp\mu}\frac{p_{\nu}}{z}\right]\nonumber \\
  &\quad \quad \times\left(\lambda_h D^\perp_{L} +\frac{k_\perp \cdot S_T}{M}D^\perp_{T}\right) \nonumber\\
  & -\frac{2}{p\cdot q}\left[c_3^q  \varepsilon^k_{\perp\mu}\frac{p_{\nu}}{z}+ic_1^q k_{\perp\mu}\frac{p_{\nu}}{z}\right]G^\perp  \nonumber\\
  & +\frac{2M}{p\cdot q}\left[c_3^q  S_{T\mu}\frac{p_{\nu}}{z}-ic_1^q \varepsilon^S_{\perp\mu}\frac{p_{\nu}}{z}\right]G_T  \nonumber\\
  & + \frac{2}{p\cdot q}\left[c_3^q  k_{\perp\mu}n_{\nu}-ic_1^q \varepsilon^k_{\perp\mu}\frac{p_{\nu}}{z}\right]\nonumber \\
  &\quad \quad \times\left(\lambda_h G^\perp_{L} +\frac{k_\perp \cdot S_T}{M}G^\perp_{T}\right).
 \label{f:W1Lt3}
\end{align}
We have provided the detailed calculation for obtaining this expression in Appendix \ref{sec:t3}.

The complete twist-3 hadronic tensor is the sum of all the twist-3 contributions, i.e, $ W_{t3\mu\nu}= W^{(0)}_{t3\mu\nu}+ W^{(1)L}_{t3\mu\nu}+\left(W^{(1)L}_{t3\nu\mu}\right)^*$. Finally, the explicit expression is given by 
\begin{align}
  z W_{t3\mu\nu} = & -\frac{2}{p\cdot q}\left[c_1^q  k_{\perp\{\mu}\bar{q}_{\nu\}}-ic_3^q \varepsilon^k_{\perp[\mu}\bar{q}_{\nu]}\right]D^\perp  \nonumber\\
  & -\frac{2M}{p\cdot q}\left[c_1^q  \varepsilon^S_{\perp\{\mu}\bar{q}_{\nu\}}+ic_3^q S_{T[\mu}\bar{q}_{\nu]}\right]D_T  \nonumber\\
  & - \frac{2}{p\cdot q}\left[c_1^q  \varepsilon^k_{\perp\{\mu}\bar{q}_{\nu\}}+ic_3^q k_{\perp[\mu}\bar{q}_{\nu]}\right]\nonumber \\
  &\quad \quad \times\left(\lambda_h D^\perp_{L} +\frac{k_\perp \cdot S_T}{M}D^\perp_{T}\right) \nonumber\\
  & +\frac{2}{p\cdot q}\left[c_3^q  \varepsilon^k_{\perp\{\mu}\bar{q}_{\nu\}}+ic_1^q k_{\perp[\mu}\bar{q}_{\nu]}\right]G^\perp  \nonumber\\
  & -\frac{2M}{p\cdot q}\left[c_3^q  S_{T\{\mu}\bar{q}_{\nu\}}-ic_1^q \varepsilon^S_{\perp[\mu}\bar{q}_{\nu]}\right]G_T  \nonumber\\
  & - \frac{2}{p\cdot q}\left[c_3^q  k_{\perp\{\mu}\bar{q}_{\nu\}}-ic_1^q \varepsilon^k_{\perp[\mu}\bar{q}_{\nu]}\right]\nonumber \\
  &\quad \quad \times\left(\lambda_h G^\perp_{L} +\frac{k_\perp \cdot S_T}{M}G^\perp_{T}\right),\label{f:Wt3}
\end{align}
where $\bar q^\mu =q^\mu -2p^\mu/z$. It can be shown that $W_{t3\mu\nu}$ satisfies the current conservation $q^\mu W_{t3\mu\nu}=q^\nu W_{t3\mu\nu}=0$.

\subsection{Differential cross section}

In the previous subsection, we obtained the complete hadronic tensor up to twist-3. Contracting it with the leptonic tensor \eqref{f:lepto}, one can obtain the differential cross section of the jet production semi-inclusive electron positron annihilation process.
We first give the form of the contraction results. For the leading twist case, we have 
\begin{align}
   & L_{\gamma\gamma}^{\mu\nu}\cdot \left[c_1^qg_{\perp\mu\nu}+ic_3^q\varepsilon_{\perp\mu\nu}\right]=-2Q^2 \left(c_1^q A(y)+\lambda_e c_3^q C(y)\right), \label{f:l13}\\
   & L_{\gamma\gamma}^{\mu\nu}\cdot \left[c_3^qg_{\perp\mu\nu}+ic_1^q\varepsilon_{\perp\mu\nu}\right]=-2Q^2 \left(c_3^q A(y)+\lambda_e c_1^q C(y)\right). \label{f:l31}
\end{align}
For the twist-3 case, we have 
\begin{align}
   & L_{\gamma\gamma}^{\mu\nu}\cdot \left[c_1^q  k_{\perp\{\mu}\bar{q}_{\nu\}}-ic_3^q \varepsilon^k_{\perp[\mu}\bar{q}_{\nu]}\right]\nonumber\\
   & =4Q^3 \left(c_1^q C(y)D(y)+\lambda_e c_3^q D(y)\right)|\vec{k}_\perp|\cos\phi, \label{f:3l13}\\
   & L_{\gamma\gamma}^{\mu\nu}\cdot \left[c_1^q  \varepsilon^S_{\perp\{\mu}\bar{q}_{\nu\}}+ic_3^q S_{T[\mu}\bar{q}_{\nu]}\right]\nonumber\\
   & =4Q^3 \left(c_1^q C(y)D(y)+\lambda_e c_3^q D(y)\right)|\vec{S}_T|\sin\phi_S. \label{f:3l31}
\end{align}
Other contractions for twist-3 terms can be obtained similarly. We do not show them here for simplicity.

For clarify, we divide the differential cross section into the leading twist part and the twist-3 part. For the leading twist one, we have 
\begin{align}
    d\tilde{\sigma}_{2}= &\frac{\alpha A_{ZZ}}{\pi Q^2}\Bigg\{\left[T^q_0(y)-\lambda_e \tilde{T}^q_0(y)\right]D_1\nonumber \\
    & + \left[T^q_1(y)-\lambda_e \tilde{T}^q_1(y)\right]\lambda_h G_{1L}\nonumber \\
    & - \left[T^q_0(y)-\lambda_e \tilde{T}^q_0(y)\right]k_{\perp M}|\vec{S}_T|\sin(\phi-\phi_S)D^\perp_{1T} \nonumber\\
    & -  \left[T^q_1(y)-\lambda_e \tilde{T}^q_1(y)\right]k_{\perp M}|\vec{S}_T|\cos(\phi-\phi_S)G^\perp_{1T} \Bigg\},\label{f:crosslead}
\end{align}
where $d\tilde{\sigma}=d\sigma/dzdyd^2k_\perp^\prime$, $k_{\perp M}=|\vec{k}_\perp|/M$, and 
\begin{align}
    & T^q_0(y) = c_1^ec_1^qA(y)-c_3^ec_3^qC(y), \\
    & \tilde{T}^q_0(y) = c_3^ec_1^qA(y)-c_1^ec_3^qC(y), \\
    & T^q_1(y) = c_1^ec_3^qA(y)-c_3^ec_1^qC(y), \\
    & \tilde{T}^q_1(y) = c_3^ec_3^qA(y)-c_1^ec_1^qC(y). 
\end{align}
The sum of the differential cross section for electromagnetic, interference, and weak terms is understood.
The twist-3 differential cross section can be expressed as
\begin{align}
     d\tilde{\sigma}_{3} =&-\frac{\alpha A_{ZZ}}{\pi Q^2}\frac{4\kappa_M}{z}\Bigg\{
     \left[T^q_2(y)-\lambda_e \tilde{T}^q_2(y)\right]k_{\perp M} \cos\phi D^\perp \nonumber\\
     & +\left[T^q_2(y)-\lambda_e \tilde{T}^q_2(y)\right]k_{\perp M}\sin\phi \lambda_h D^\perp_L \nonumber \\
     & +\left[T^q_2(y)-\lambda_e \tilde{T}^q_2(y)\right]|\vec{S}_T|\sin\phi_S\left(D_T-\frac{k_{\perp M}^2}{2}D^\perp_T\right) \nonumber\\
     &- \left[T^q_2(y)-\lambda_e \tilde{T}^q_2(y)\right] |\vec{S}_T|\sin(2\phi-\phi_S)\frac{k_{\perp M}^2}{2}D^\perp_T \nonumber\\
     &- \left[T^q_3(y)-\lambda_e \tilde{T}^q_3(y)\right]k_{\perp M} \sin\phi G^\perp \nonumber\\
     & +\left[T^q_3(y)-\lambda_e \tilde{T}^q_3(y)\right]k_{\perp M}\cos\phi \lambda_h G^\perp_L  \nonumber\\
     & +\left[T^q_3(y)-\lambda_e \tilde{T}^q_3(y)\right]|\vec{S}_T|\cos\phi_S\left(G_T-\frac{k_{\perp M}^2}{2}G^\perp_T\right) \nonumber\\
     &- \left[T^q_3(y)-\lambda_e \tilde{T}^q_3(y)\right] |\vec{S}_T|\cos(2\phi-\phi_S)\frac{k_{\perp M}^2}{2}G^\perp_T
     \Bigg\}, \label{f:crosst3}
\end{align}
where $\kappa_M=M/Q$ and $T$ functions are defined as 
\begin{align}
    & T^q_2(y) = c_1^ec_1^qC(y)D(y)-c_3^ec_3^qD(y), \\
    & \tilde{T}^q_2(y) = c_3^ec_1^qC(y)D(y)-c_1^ec_3^qD(y),  \\
    & T^q_3(y) =c_1^ec_3^qC(y)D(y)-c_3^ec_1^qD(y),  \\
    & \tilde{T}^q_3(y) = c_3^ec_3^qC(y)D(y)-c_1^ec_1^qD(y).
\end{align}

Up to now, we have presented the results of the differential cross section using the weak interaction term as an example. To obtain the results for the electromagnetic term, we need to set $c_3^{e/q} = 0$ and $c_1^{e/q} = 1$. For the interference term, we need to set $c_3^{e/q}=c_A^{e/q}$ and $c_1^{e/q}=c_V^{e/q}$.
The kinematic factors are also different.
To make it transparent, we can get the electromagnetic and interference cross sections by replacing the parameters in the weak interaction cross section according to Table.~\ref{tab:replacing}

\begin{table}
\renewcommand\arraystretch{1.5}
\begin{tabular}{cccc}
\hline \hline
~~~~  & $A_r$  & $L^{\mu\nu}_r$  & $W^{\mu\nu}_r$  \\ \hline 
~ $ZZ$~ & $A_{ZZ}$ & $c_1^e,~c_3^e$ & $c_1^q,~c_3^q$  \\
$\gamma Z$  & ~$A_{ZZ}\to A_{\gamma Z}$ ~& ~~$c_1^e\to c_V^e,~c_3^e\to c_A^e$~~ &~~ $c_1^q\to c_V^q,~c_3^q\to c_A^q$~~ \\
$\gamma\gamma$ & $A_{ZZ}\to A_{\gamma \gamma}$  & $c_1^e\to 1,~c_3^e\to 0$  & $c_1^q\to 1,~c_3^q\to 0$ \\ \hline \hline
\end{tabular}
\caption{Relations of kinematic factors between weak, electromagnetic, and interference interactions.}
\label{tab:replacing}
\end{table}

\section{Numerical estimates of the transverse momentum asymmetry}\label{sec:results}

As introduced in Sec. \ref{sec:frame}, the transverse momentum asymmetry is defined through the difference of the transverse momenta, see Eqs. \eqref{f:axk} and \eqref{f:ayk}. Note that the transverse momentum of the fragmenting quark lies in the $x$-$y$ plane, and it can be decomposed as
\begin{align}
 & k_\perp^{x}=|\vec{k}_\perp| \cos\phi, \\
 & k_\perp^{y}=|\vec{k}_\perp| \sin\phi.
\end{align}
We notice that 
\begin{align}
  & k_\perp^x(x>0)\to\int_{-\pi/2}^{\pi/2}|\vec{k}_\perp| \cos\phi d\phi, \\
  & k_\perp^{x} (x<0)\to\int_{\pi/2}^{3\pi/2}|\vec{k}_\perp| \cos\phi d\phi.
\end{align}
Therefore, $k_\perp^x(x>0)$ and $k_\perp^x(<0)$ can be related to the differential cross section via the following equations:
\begin{align}
   k_\perp^x(x>0) &\to\int_{x>0}\vec{k}_\perp d^2k_\perp \frac{d\sigma}{d^2k_\perp}=\int_{x>0}k_\perp^2 \cos \phi dk_\perp d\phi \frac{d\sigma}{d^2k_\perp} \nonumber \\
  &=\int^{\pi/2}_{-\pi/2}k_\perp^2 \cos \phi dk_\perp d\phi \frac{d\sigma}{d^2k_\perp dz dy}dz dy, \\
  k_\perp^x(x>0) &\to\int_{x>0}\vec{k}_\perp d^2k_\perp \frac{d\sigma}{d^2k_\perp}=\int_{x>0}k_\perp^2 \cos \phi dk_\perp d\phi \frac{d\sigma}{d^2k_\perp} \nonumber \\
  &=\int^{3\pi/2}_{\pi/2}k_\perp^2 \cos \phi dk_\perp d\phi \frac{d\sigma}{d^2k_\perp dz dy}dz dy,  
\end{align}
where $\frac{d\sigma}{d^2k_\perp dz dy}$ is the complete differential cross section shown in Eqs. \eqref{f:crosslead} and \eqref{f:crosst3}. For the production of the unpolarized hadron, we therefore have 
\begin{align}
  k_\perp^x(x>0)&= \frac{2\alpha A_{ZZ}}{ Q^2}\left[T^q_0(y)-\lambda_e \tilde{T}^q_0(y)\right]k_\perp^2 D_1 \nonumber \\
  &-\frac{\alpha A_{ZZ}}{Q^2}\frac{2}{zQ}\left[T^q_2(y)-\lambda_e \tilde{T}^q_2(y)\right] k_{\perp}^3 D^\perp, \label{f:kperpp}\\
  k_\perp^{x} (x<0)&=\frac{2\alpha A_{ZZ}}{ Q^2}\left[T^q_0(y)-\lambda_e \tilde{T}^q_0(y)\right]k_\perp^2 D_1 \nonumber \\
  &+\frac{\alpha A_{ZZ}}{Q^2}\frac{2}{zQ}\left[T^q_2(y)-\lambda_e \tilde{T}^q_2(y)\right] k_{\perp}^3 D^\perp. \label{f:kperpm}
\end{align}


According to the definitions shown in Eqs. \eqref{f:axk} and \eqref{f:ayk} and the results shown above, four kinds of asymmetries are expressed as
\begin{align}
 & A_{U}^x = -\frac{\kappa_M k_{\perp M}}{z}\frac{A_{ZZ}}{A_{ZZ}} \frac{\left[T^q_2(y)-\lambda_e \tilde{T}^q_2(y)\right]}{\left[T^q_0(y)-\lambda_e \tilde{T}^q_0(y)\right]}\frac{D^\perp}{D_1}, \label{f:auux} \\
 & A_{U}^y = +\frac{\kappa_M k_{\perp M}}{z}\frac{A_{ZZ}}{A_{ZZ}} \frac{\left[T^q_3(y)-\lambda_e \tilde{T}^q_3(y)\right]}{\left[T^q_0(y)-\lambda_e \tilde{T}^q_0(y)\right]}\frac{G^\perp}{D_1}, \label{f:auuy} \\
 & A_{L}^x = - \frac{\kappa_M k_{\perp M}}{z}\frac{A_{ZZ}}{A_{ZZ}} \frac{\left[T^q_3(y)-\lambda_e \tilde{T}^q_3(y)\right]}{\left[T^q_0(y)-\lambda_e \tilde{T}^q_0(y)\right]}\frac{G_L^\perp}{D_1}, \label{f:aulx} \\
 & A_{L}^y = -\frac{\kappa_M k_{\perp M}}{z}\frac{A_{ZZ}}{A_{ZZ}} \frac{\left[T^q_2(y)-\lambda_e \tilde{T}^q_2(y)\right]}{\left[T^q_0(y)-\lambda_e \tilde{T}^q_0(y)\right]}\frac{D_L^\perp}{D_1}, \label{f:auly} 
\end{align}
where the subscripts $U$ and $L$ denote unpolarized and the longitudinally polarized hadrons, respectively. We note that these transverse momentum asymmetries are all twist-3 measurable quantities. We note again that only weak interaction results are shown in Eqs. \eqref{f:auux}--\eqref{f:auly}. For the complete results, electromagnetic and interference terms should be included.

If only the electromagnetic interaction is considered, the resulting asymmetries are given by
\begin{align}
 & A_{U}^x = -\frac{\kappa_M k_{\perp M}}{z}\frac{e_q^2}{e_q^2} \frac{C(y)D(y)}{A(y)}\frac{D^\perp}{D_1}, \label{f:auuxem} \\
 & A_{U}^y = -\frac{\kappa_M k_{\perp M}}{z}\frac{e_q^2}{e_q^2} \frac{\lambda_eD(y)}{A(y)}\frac{G^\perp}{D_1}, \label{f:auuyem} \\
 & A_{L}^x = + \frac{\kappa_M k_{\perp M}}{z}\frac{e_q^2}{e_q^2} \frac{\lambda_eD(y)}{A(y)}\frac{G_L^\perp}{D_1}, \label{f:aulxem} \\
 & A_{L}^y = -\frac{\kappa_M k_{\perp M}}{z}\frac{e_q^2}{e_q^2} \frac{C(y)D(y)}{A(y)}\frac{D_L^\perp}{D_1}. \label{f:aulyem} 
\end{align}

\begin{figure}
\centering
\includegraphics[width= 0.8\linewidth]{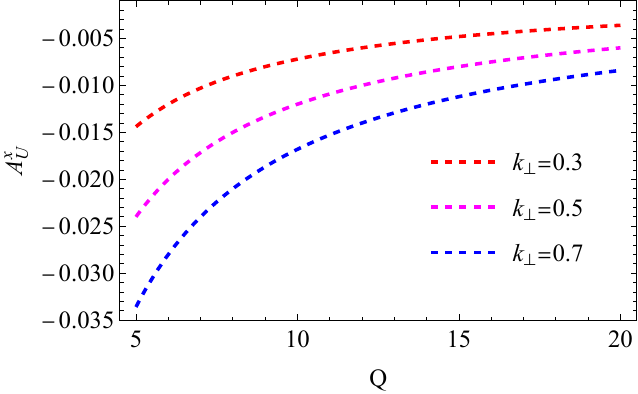}
\caption{Numerical estimates of transverse momentum asymmetry $A_{U}^x$ with respect to $Q$. In this case $z=0.2$ and $y=0.4$.}
\label{fig:Auq}
\end{figure}

\begin{figure}
\centering
\includegraphics[width= 0.8\linewidth]{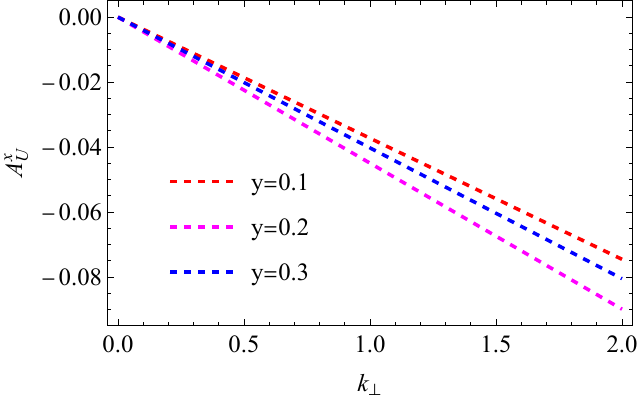}
\caption{Numerical estimates of transverse momentum asymmetry $A_{U}^x$ with respect to $k_\perp$. In this case $z=0.2$ and $Q=10$ GeV.}
\label{fig:Auk}
\end{figure}

\begin{figure}
\centering
\includegraphics[width= 0.8\linewidth]{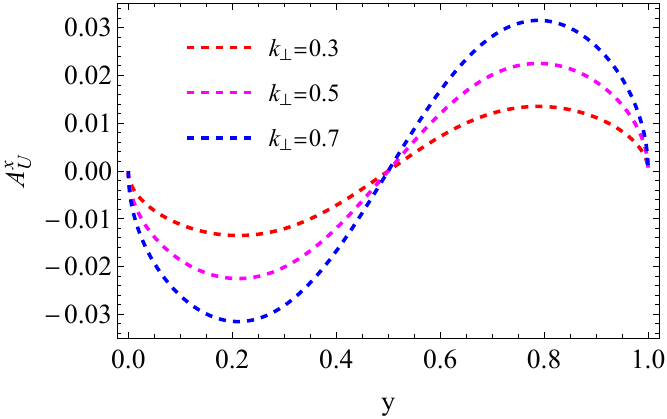}
\caption{
Numerical estimates of transverse momentum asymmetry $A_{U}^x$ with respect to $y$. In this case $z=0.2$ and $Q=10$ GeV.}
\label{fig:Auy}
\end{figure}

\begin{figure}
\centering
\includegraphics[width= 0.8\linewidth]{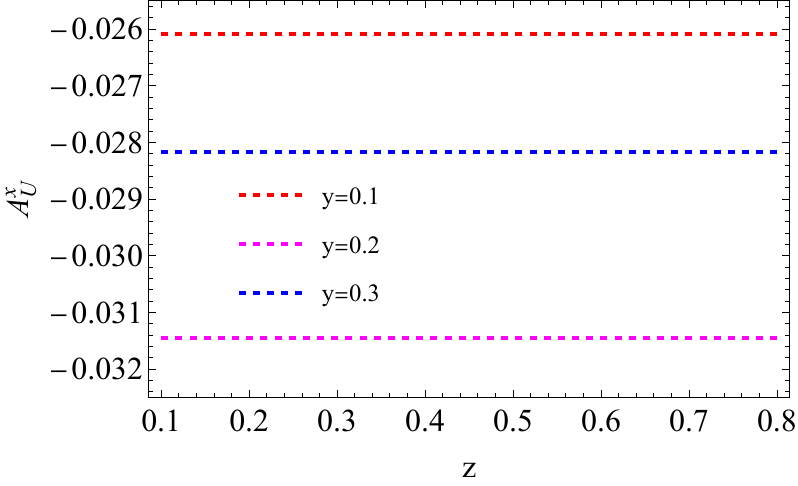}
\caption{Numerical estimates of transverse momentum asymmetry $A_{U}^x$ with respect to $z$. In this case $Q=10$ GeV and $k_\perp=0.7$ GeV.}
\label{fig:Auz}
\end{figure}

To have an intuitive impression of the transverse momentum asymmetries introduced before, we present the numerical estimates of $A_{U}^x$ in Fig. \ref{fig:Auq}, Fig. \ref{fig:Auk}, Fig. \ref{fig:Auy}, and Fig. \ref{fig:Auz} with respect to $Q$, $k_\perp$, $y$, and $z$, respectively. We take the Gaussian ansatz for the TMD FFs, i.e.,
\begin{align}
 & D_1(z, k_\perp) =\frac{1}{\pi {\Delta^\prime}^2} D_1(z) e^{-\vec k_\perp^2/{\Delta^\prime}^2}, \\
 & D^\perp(z, k_\perp) =\frac{z}{2\pi \Delta^2}D_1(z) e^{-\vec k_\perp^2/\Delta^2}, \label{f:Dperp}
\end{align}
where $D_1(z)$ are taken from NPC23 \cite{Gao:2024nkz,Gao:2024dbv} for the $\pi^+$ meson.
We only take the light flavors ($u, d, s$) into account.
To determine $D^\perp(z, k_\perp)$, we have used the Wandzura-Wilczek approximation (neglecting the quark-gluon-quark correlation function, $g=0$) \cite{Wandzura:1977qf} and the massless QCD equation of motion $\slashed D\psi=0$.  The mean squared transverse momenta of TMD FFs are taken as ${\Delta^\prime}^2=0.17$ GeV$^2$ and $\Delta^2=0.17$ GeV$^2$ for all flavors
\cite{Anselmino:2005nn,Signori:2013mda,Anselmino:2013lza,Cammarota:2020qcw,Bacchetta:2022awv,Bacchetta:2024qre}.

Under the condition of Gaussian ansatz for the TMD FFs, numerical estimates lead to three observations illustrated below.
\begin{itemize}

 \item From Fig. \ref{fig:Auq}, we notice that the absolute values of the transverse momentum asymmetry, $A_{U}^x$, decreases with increasing energy $Q$ due to the power suppression of the twist-3 effects. Nevertheless, the absolute values of the asymmetry increases with respect to the transverse momentum $k_\perp$. To be more precise, $|A_{U}^x|$ is proportional to $k_\perp$, as illustrated in Fig. \ref{fig:Auk}.

 \item From Fig. \ref{fig:Auy}, we notice that $A_{U}^x$ exhibits periodic characteristics. It is also fixed at $y=0,\, 0.5,\, 1$, 
 \begin{align}
    A_{U}^x(y=0)=A_{U}^x(y=0.5)=A_{U}^x(y=1)=0. 
 \end{align}
   Note that $y=p\cdot l_1/p\cdot q \sim (1+\cos\theta)/2$, where $\theta$ is the angle between the positron ($l_2$) and the observed hadron ($p$). 
   This indicates that, according to our conventions, the asymmetry is positive when the observed hadron is in the electron-beam hemisphere and negative when it is in the positron-beam hemisphere.
   
 \item Figure \ref{fig:Auz} shows that asymmetry $A_U^x$ is not sensitive to fraction $z$. From Eqs. \eqref{f:axk} and \eqref{f:ayk}, we notice that the asymmetry is defined through the difference of the transverse momenta, and it should therefore be independent of the longitudinal momentum fraction $z$. Certainly, this conclusion strongly depends on the parametrized form of $D^\perp$ in Eq. \eqref{f:Dperp},
 which implies that $D^\perp$ may be proportional to $zD_1$.

\end{itemize}

We also check that asymmetry $A_{U}^x$ is not sensitive to the mean squared transverse momenta of TMD FFs. Furthermore, the use of NNFF10 \cite{Bertone:2017tyb} for $\pi^+$, MAPFF10 \cite{AbdulKhalek:2022laj} for $\pi^+$, and NPC23 \cite{Gao:2024nkz,Gao:2024dbv} for $K^+$ yields consistent results. 
These findings suggest that the transverse momentum asymmetry is very likely a universal physical quantity that appears to be independent of the identity of the observed hadronic state.
It reveals the transverse momentum effects and does not depend on the longitudinal momentum fraction. 


\section{Summary}\label{sec:summary}


Hadronization, being a nonperturbative QCD process, cannot be calculated from first principles and is typically described through phenomenological models. Within the framework of the QCD parton model, one can study the hadronization process through the behaviors of the single hadron and the double hadron or through FFs and diFFs. However, higher-twist FFs present significant challenges for determination due to power suppression. We notice that the transverse momentum of the produced jet or the intrinsic quark in the semi-inclusive electron positron annihilation process exhibits twist-3 effects. Under this circumstance, we propose a novel observable: the transverse momentum asymmetry of the fragmenting quark with respect to the observed hadron direction. This asymmetry is defined as the difference between the transverse momentum components of the fragmenting quark relative to the observed hadron, especially in the $x$ and $y$ directions in the center-of-mass frame of leptons. It is a twist-3 effect and provides direct access to twist-3 TMD FFs. 

To investigate the transverse momentum effects, we develop a general theoretical framework up to twist-3 at leading order for the jet production semi-inclusive annihilation process in this paper. Higher-order gluon radiations are not considered.
We derive the differential cross section by incorporating both electromagnetic and weak interactions and express the resulting asymmetries in terms of twist-3 TMD FFs. 
Taking the Gaussian ansatz for the TMD FFs, we present numerical estimates of the asymmetry $A_U^x$.
The results show that the absolute values of the asymmetry decreases with increasing center-of-mass energy $Q$ but increases linearly with respect to the transverse momentum $k_\perp$.
Furthermore, the asymmetry exhibits periodic behavior with respect to fraction $y$ and does not depend on the longitudinal momentum fraction $z$. 

The transverse momentum asymmetry serves as a measurable and sensitive probe for studying jet dynamics in the annihilation process. It also offers valuable insights into the relative distributions between the observed hadron and the fragmenting quark during hadronization. This provides a new approach to understand the hadronization mechanism. Based on the definitions and calculations, we notice that the asymmetry is very likely to be a universal physical quantity that is insensitive to the type of hadron and the longitudinal momentum fraction.
Furthermore, it provides a pathway ($D^\perp \sim zD_1$) to determine twist-3 TMD FFs from future experimental data. 

\section*{Acknowledgments}
This work was supported by the National Natural Science Foundation of China (Grants No. 12405103, and No. 12305106), and the Youth Innovation Technology Project of Higher School in Shandong Province (2023KJ146).

\begin{appendix}

\section{$\Xi^{(1)}_\rho$ CONTRIBUTIONS} \label{sec:t3}

 It is worthwhile to illustrate that the twist-3 fragmentation functions given in Eqs. \eqref{f:xi0even}, \eqref{f:xi0odd} and \eqref{f:xi1even3}, \eqref{f:xi1odd3} are not independent. Their relationships can be derived by applying the QCD equation of motion, $\slashed{D} \psi = 0$. Eliminating the longitudinal components of the correlators yields the relationships among the transverse components.
 For the two transverse components $\Xi_{\perp}^{(0)\rho}$ and $\tilde\Xi_{\perp}^{(0)\rho}$, we have \cite{Yang:2017sxz}
\begin{align}
  k^+\Xi_{\perp}^{(0)\rho}&=-g_\perp^{\rho\sigma} \mathrm{Re}\Xi^{(1)}_{\sigma +}-\varepsilon^{\rho\sigma}_{\perp }\mathrm{ Im} \tilde \Xi^{(1)}_{\sigma +},\label{f:perpeom}\\
  k^+\tilde \Xi_{\perp}^{(0)\rho}&=-g_\perp^{\rho\sigma} \mathrm{Re}\tilde \Xi^{(1)}_{\sigma +}-\varepsilon^{\rho\sigma}_{\perp }\mathrm{ Im} \Xi^{(1)}_{\sigma +}. \label{f:tperpeom}
\end{align}
Equations (\ref{f:perpeom}) and (\ref{f:tperpeom}) lead to a set of relationships between twist-3 TMD FFs
\begin{align}
  & D^\perp - iG^\perp =-z(D_d^\perp-G_d^\perp), \label{f:adp}\\
  & D_{T} - iG_{T} =-z(D_{dT}-G_{dT}), \\
  & D^\perp_L - iG^\perp_L =-z(D^\perp_{dL}-G^\perp_{dL}), \\
  & D^\perp_{T} - iG^\perp_{T} =-z(D^\perp_{dT}-G^\perp_{dT}).
\end{align}
Substituting Eqs. (\ref{f:xi1even3}) and (\ref{f:xi1odd3}) into Eq. (\ref{f:tW1L}), we obtain
\begin{align}
  z W^{(1,L)}_{t3\mu\nu} = & -\frac{2}{p\cdot q}\left[c_1^q k_{\perp\mu}p_{\nu} -ic_3^q \varepsilon^k_{\perp\mu}p_{\nu}\right]D_d^\perp  \nonumber\\
  & -\frac{2M}{p\cdot q}\left[c_1^q  \varepsilon^S_{\perp\mu}p_{\nu}+ic_3^q S_{T\mu}p_{\nu}\right]D_{dT}  \nonumber\\
  & - \frac{2}{p\cdot q}\left[c_1^q  \varepsilon^k_{\perp\mu}p_{\nu}+ic_3^q k_{\perp\mu}p_{\nu}\right]\nonumber \\
  &\quad \quad \times\left(\lambda_h D^\perp_{dL} +\frac{k_\perp \cdot S_T}{M}D^\perp_{dT}\right) \nonumber\\
  & +\frac{2}{p\cdot q}\left[c_3^q  \varepsilon^k_{\perp\mu}p_{\nu}-ic_1^q k_{\perp\mu}p_{\nu}\right]G_d^\perp  \nonumber\\
  & +\frac{2M}{p\cdot q}\left[c_3^q  S_{T\mu}p_{\nu}+ic_1^q \varepsilon^S_{\perp\mu}p_{\nu}\right]G_{dT}  \nonumber\\
  & + \frac{2}{p\cdot q}\left[c_3^q  k_{\perp\mu}n_{\nu}+ic_1^q \varepsilon^k_{\perp\mu}p_{\nu}\right]\nonumber \\
  &\quad \quad \times\left(\lambda_h G^\perp_{dL} +\frac{k_\perp \cdot S_T}{M}G^\perp_{dT}\right).
 \label{f:W1Lt3d}
\end{align}
Subsequently, we employ the relationships given in Eqs. \eqref{f:adp}--\eqref{f:W1Lt3d} to express the functions $D_d$ and $G_d$ in terms of $D$ and $G$. This substitution leads us to the final result presented in Eq. \eqref{f:W1Lt3}.

\end{appendix}

\end{document}